\documentclass[12pt]{iopart}

\usepackage{times}

\makeatletter
\@addtoreset{equation}{section}
\makeatother
\renewcommand{\theequation}{\thesection.\arabic{equation}}
\begin{document}

{\title{\Large {\bf Spin 1/2 bosons etc. in a theory with Lorentz violation}}}

\author{Roland E. Allen\\
{\normalfont Center for Theoretical Physics, Texas A\&M University}}

\begin{abstract}
An action with unconventional supersymmetry was introduced in an earlier 
paper. Here it is shown that this action leads to standard physics 
for fermions and gauge bosons at low energy, but to testable 
extensions of standard physics for fermions at high energy 
and for fundamental bosons which have not yet been observed. For 
example, the Lorentz-violating equation of motion for these bosons 
implies that they have spin 1/2.
\end{abstract}

\section{Introduction}

The following Euclidean action was postulated in an earlier 
paper~\cite{allen1}:
\begin{equation}
S=\int d^{D}x\left[ \frac{1}{2m}\partial ^{M}\Psi ^{\dagger }\partial
_{M}\Psi -\mu \Psi ^{\dagger }\Psi +\frac{1}{2}b\left( \Psi ^{\dagger }\Psi
\right) ^{2}\right]
\end{equation}
with 
\begin{equation}
\Psi =\left( 
\begin{array}{c}
z_{1} \\ 
z_{2} \\ 
\vdots \\ 
z_{N}
\end{array}
\right) \qquad ,\qquad z=\left( 
\begin{array}{c}
z_{b} \\ 
z_{f}
\end{array}
\right) .
\end{equation}
This action has ``natural supersymmetry'', in the sense that the initial
bosonic fields $z_{b}$ and fermionic fields $z_{f}$ are treated in exactly
the same way. The only difference is that the $z_{b}$ are ordinary complex
numbers whereas the $z_{f}$ are anticommuting Grassmann numbers. (Here, as in 
Ref. 1, ``supersymmetry'' is taken to have its general definition~\cite{parisi, 
efetov}: An action is supersymmetric if it is invariant under a
transformation which converts fermions to bosons and vice-versa.) It was 
found in Ref. 1 that standard physics can emerge from (1.1) at energies that
are far below the Planck scale, provided that specific kinds of topological
defects are included in the theory. For example, one can obtain an $SO(10)$
grand-unified theory, containing both the Standard Model and a natural
mechanism for small neutrino masses [4-15].

In the present paper, it will be shown that the theory predicts testable 
extensions of standard physics, both for fermions at high energy and for 
fundamental bosons which have not yet been observed.

\section{Canonical Quantization in Lorentzian Spacetime}

Path-integral quantization can ordinarily be replaced by canonical 
quantization, or vice-versa~\cite{weinberg}, through a procedure 
that is similar to that for a single particle. In the present 
theory, whether this can be done consistently is a nontrivial issue, 
because the resulting field theory 
has some very unconventional features. These will be discussed in 
Section 4, but in the present section it will simply be assumed that 
one can define quantized fields $\hat{\Psi }$ etc. in the usual way~[16-24]. 

After a change from path-integral to canonical quantization, and an inverse 
Wick rotation from Euclidean to Lorentzian time (with $S_{L}=iS$), the 
action (1.1) becomes 
\begin{equation}
\hat{S}_{L}=-\int d^{D}x\left[ \frac{1}{2m}\eta ^{MN}\partial _{M}
\hat{\Psi }_{L}^{\dagger }\partial _{N}\hat{\Psi }_{L}-\mu \hat{
\Psi }_{L}^{\dagger }\hat{\Psi }_{L}+\frac{1}{2}b\left( \hat{\Psi }
_{L}^{\dagger }\hat{\Psi }_{L}\right) ^{2}\right]
\end{equation}
where $\eta ^{MN}=diag(-1,1,...,1)$. This notation is rather awkward, 
however, so for the remainder of the paper we will let
\begin{equation}
\hat{S}_{L}\rightarrow S,\;\hat{\Psi }_{L}\rightarrow \Psi
\end{equation}
with the understanding that these are now quantized operators in 
Lorentzian spacetime. It is also understood that raising and 
lowering of indices is now done with the Minkowski metric tensor: 
\begin{equation}
A^{\mu }B_{\mu }=\eta ^{\mu \nu }A_{\mu }B_{\nu }\quad 
\mbox{or in $D$ 
dimensions}\quad A^{M}B_{M}=\eta ^{MN}A_{M}B_{N}.
\end{equation}
Later in this paper we will introduce the metric tensor associated with
gravity and general coordinate transformations. To avoid confusion, this
metric tensor $g_{\mu \nu }$ will always be shown explicitly, and simple
raising and lowering of indices will always have the interpretation (2.3).

With the above change of notation, and after an integration by parts, 
(2.1) becomes
\begin{equation}
S=-\int d^{D}x\left[ -\frac{1}{2m}\Psi ^{\dagger }\partial ^{M}\partial
_{M}\Psi -\mu \Psi ^{\dagger }\Psi +\frac{1}{2}b\left( \Psi ^{\dagger }\Psi
\right) ^{2}\right] .
\end{equation}
The resulting equation of motion is 
\begin{equation}
\hspace{-1.5cm}
\left[ -\frac{1}{2m}\partial ^{M}\partial _{M}-\mu +V_{vac}+b\Delta \left(
\Psi ^{\dagger }\Psi \right) \right] \Psi =0\quad ,\quad
V_{vac}=b\left\langle \Psi ^{\dagger }\Psi \right\rangle _{vac}
\end{equation}
where $\left\langle \cdot \cdot \cdot \right\rangle _{vac}$ represents a
vacuum expectation value, and 
\begin{equation}
\Psi ^{\dagger }\Psi =\left\langle \Psi ^{\dagger }\Psi \right\rangle
_{vac}+\Delta \left( \Psi ^{\dagger }\Psi \right) .
\end{equation}
For the remainder of this section, we will consider either the vacuum or a
noninteracting free field in the vacuum. We then have 
\begin{equation}
\hspace{-1.5cm}
\left( -\frac{1}{2m}\partial ^{M}\partial _{M}-\mu 
+V_{vac}\right) \Psi
_{b}=0\quad ,\quad \left( -\frac{1}{2m}\partial ^{M}\partial _{M}-\mu
+V_{vac}\right) \Psi _{f}=0.
\end{equation}

It will be assumed that the physical vacuum contains a condensate whose
order parameter 
\begin{equation}
\Psi _{cond}=\left\langle \Psi_{b} \right\rangle _{vac}
\end{equation}
has the form 
\begin{eqnarray}
\Psi _{cond}&=&U\,n_{cond}^{1/2}\eta _{0} \\
U^{\dagger }U&=&\eta _{0}^{\dagger }\eta _{0}=1 .
\end{eqnarray}
(As discussed in the next section, $\Psi _{cond}$ is dominantly due to a GUT
field that condenses in the very early universe. In the present theory, it
is not static, but instead exhibits rotations in space and time that are
described by $U$.) It will also be assumed that the order parameter can be
written in the form 
\begin{equation}
\Psi _{cond}=\Psi _{ext}\left( x^{\mu }\right) \,\Psi _{int}\left(
x^{m},x^{\mu }\right)
\end{equation}
\begin{eqnarray}
\Psi _{ext}\left( x^{\mu }\right) = U_{ext}\left( x^{\mu }\right)
\,n_{ext}^{1/2}\left( x^{\mu }\right) \eta _{ext} \\
\Psi _{int} = U_{int}\,n_{int}^{1/2}\eta _{int}
\end{eqnarray}
where $\eta _{ext}$ and $\eta _{int}$ are constant vectors, and 
the quantities in the lower equation can depend on $x^{\mu }$ as well
as $x^{m}$. Let us define external and internal ``superfluid velocities'' by 
\begin{equation}
mv_{M}=-iU^{-1}\partial _{M}U
\end{equation}
or 
\begin{eqnarray}
mv_{\mu } &=&-iU_{ext}^{-1}\partial _{\mu }U_{ext}-iU_{int}^{-1}\partial
_{\mu }U_{int} \\
mv_{m}~ &=&-iU_{int}^{-1}\partial _{m}U_{int}.
\end{eqnarray}
The fact that $U$ is unitary implies that $\partial _{M}U^{\dagger
}U=-U^{\dagger }\partial _{M}U$ with $U^{\dagger }=U^{-1}$, or 
\begin{equation}
mv_{M}=i\partial _{M}U^{\dagger }U
\end{equation}
so that 
\begin{equation}
v_{M}^{\dagger }=v_{M} \, .
\end{equation}

In this section we will assume that 
\begin{equation}
\partial _{\mu }U_{int}=0
\end{equation}
in which case there are separate equations of motion for external and
internal spacetime: 
\begin{equation}
\left( -\frac{1}{2m}\partial ^{\mu }\partial _{\mu }-\mu _{ext}\right) \Psi
_{ext}=0
\end{equation}
\begin{equation}
\left( -\frac{1}{2m}\partial ^{m}\partial _{m}-\mu _{int}+V_{vac}\right)
\Psi _{int}=0
\end{equation}
with $\mu _{int}=\mu -\mu _{ext}$. The quantities $V_{vac}$, $\mu _{int}$,
and $\Psi _{int}$ are allowed to have a slow parametric dependence on 
$x^{\mu }$, as long as $\partial ^{\mu }\partial _{\mu }\Psi _{int}$ is
negligible.

When (2.12), (2.15), and (2.19) are used in (2.20), we obtain 
\begin{equation}
\hspace{-1.5cm}
\eta _{ext}^{\dagger }n_{ext}^{1/2}\left[ 
\left( \frac{1}{2}mv^{\mu }v_{\mu
}-\frac{1}{2m}\partial ^{\mu }\partial _{\mu }-\mu _{ext}\right) -i\left( 
\frac{1}{2}\partial ^{\mu }v_{\mu }+v^{\mu }\partial _{\mu }\right) \right]
n_{ext}^{1/2}\eta _{ext}=0
\end{equation}
and its Hermitian conjugate 
\begin{equation}
\hspace{-1.5cm}
\eta _{ext}^{\dagger }n_{ext}^{1/2}\left[ 
\left( \frac{1}{2}mv^{\mu }v_{\mu
}-\frac{1}{2m}\partial ^{\mu }\partial _{\mu }-\mu _{ext}\right) +i\left( 
\frac{1}{2}\partial ^{\mu }v_{\mu }+v^{\mu }\partial _{\mu }\right) \right]
n_{ext}^{1/2}\eta _{ext}=0.
\end{equation}
Subtraction gives the equation of continuity 
\begin{equation}
\partial _{\mu }j_{ext}^{\mu }=0~\quad ,\quad j_{ext}^{\mu }=\eta
_{ext}^{\dagger }\,n_{ext}v^{\mu }\eta _{ext}
\end{equation}
and addition gives the Bernoulli equation 
\begin{equation}
\frac{1}{2}m\bar{v}_{ext}^{2}+P_{ext}=\mu _{ext}
\end{equation}
where
\begin{equation}
\bar{v}_{ext}^{2}=\eta _{ext}^{\dagger }\,v^{\mu }v_{\mu }\,\eta _{ext}
\end{equation}
\begin{equation}
P_{ext}=-\frac{1}{2m}n_{ext}^{-1/2}\partial ^{\mu }\partial _{\mu
}n_{ext}^{1/2}.
\end{equation}

In the present theory, the order parameter in external spacetime, $\Psi
_{ext}$, has the symmetry group $U(1)\times SU(2)$. The ``superfluid
velocity'' in external spacetime, $v_{\mu }$, can then be written in terms of
the identity matrix $\sigma ^{0}$ and Pauli matrices $\sigma ^{a}$ :
\begin{equation}
v^{\mu }=v_{\alpha }^{\mu }\sigma ^{\alpha }~~~,~~~\mu ,\alpha =0,1,2,3.
\end{equation}
\noindent It is assumed that the basic texture of the order parameter is
such that 
\begin{equation}
v_{k}^{0}=v_{0}^{a}=0~~,~~~k,a=1,2,3
\end{equation}
to a good approximation, yielding the simplification
\begin{equation}
{\frac{1}{2}}mv^{\alpha \mu }v_{\mu }^{\alpha }+P_{ext}=\mu _{ext}.
\end{equation}
Let 
\begin{equation}
\Delta \Psi _{b}=\Psi _{b}-\Psi _{cond}
\end{equation}
and let $\Psi _{a}$ represent either the bosonic field $\Delta \Psi _{b}$ or
the fermionic field $\Psi _{f}$. If we start with the case of a free field,
which interacts only with the condensate and other vacuum fields, (2.4)
gives 
\begin{equation}
S_{a}=-\int d^{D}x\,\Psi _{a}^{\dagger }\left( -\frac{1}{2m}\partial
^{M}\partial _{M}-\mu +V_{vac}\right) \Psi _{a}.
\end{equation}
Since $\Psi _{a}$ satisfies a linear equation involving a Hermitian
operator, it can be written in the form 
\begin{equation}
\Psi _{a}\left( x^{\mu },x^{m}\right) =\widetilde{\psi }_{a}^{r}\left(
x^{\mu }\right) \psi _{r}^{int}\left( x^{m}\right)
\end{equation}
with a summation implied over repeated indices, as usual. The 
$\widetilde{\psi }_{a}^{r}$ 
are field operators and the $\psi _{r}^{int}$ are a complete set
of basis functions in the internal space, which are required to be
orthonormal,
\begin{equation}
\int d^{D-4}x\,\psi _{r}^{int\dagger }\left( x^{m}\right) \psi _{r^{\prime
}}^{int}\left( x^{m}\right) =\delta _{rr^{\prime }} \, ,
\end{equation}
and to satisfy the internal equation of motion 
\begin{equation}
\left( -\frac{1}{2m}\partial ^{m}\partial _{m}-\mu _{int}+V_{vac}\right)
\psi _{r}^{int}\left( x^{m}\right) =\varepsilon _{r}\psi _{r}^{int}\left(
x^{m}\right) .
\end{equation}
(The $\psi _{r}^{int}$ are allowed to have a slow parametric dependence on 
$x^{\mu }$, as long as $\partial ^{\mu }\partial _{\mu }\psi _{r}^{int}$ is
negligible.) As usual, only the zero modes with $\varepsilon _{r}=0$ will be
kept, since the higher energies involve nodes in the internal space and are
comparable to $m_{P}$. When (2.33)-(2.35) are used in (2.32), the result is 
\begin{equation}
S_{a}=-\int d^{4}x\,\widetilde{\psi }_{a}^{\dagger }\left( -\frac{1}{2m}
\partial ^{\mu }\partial _{\mu }-\mu _{ext}\right) \widetilde{\psi }_{a}
\end{equation}
where $\widetilde{\psi }_{a}$ is the vector with components $\widetilde{\psi 
}_{a}^{r}$.

Let $\widetilde{\psi }_{a}$ be rewritten in the form 
\begin{equation}
\widetilde{\psi }_{a}\left( x^{\mu }\right) =U_{ext}\left( x^{\mu }\right)
\psi _{a}\left( x^{\mu }\right) .
\end{equation}
(The $2\times 2$ matrix $U_{ext}$ multiplies each of the 2-component
operators $\widetilde{\psi }_{a}^{r}$.) Here $\psi _{a}$ has a simple
interpretation: It is the field seen by an observer in the frame of
reference that is moving with the condensate. In the present theory, the GUT
condensate $\Psi _{cond}$ forms in the very early universe, and the other
bosonic and fermionic fields $\Psi _{a}$ are subsequently born into it. It
is therefore natural to view them from the perspective of the condensate.

Equation (2.37) is, in fact, exactly analogous to rewriting the wavefunction
of a particle in an ordinary superfluid moving with velocity $v_{s}$: $\psi
_{p}^{\prime }\left( x\right) =\exp \left( iv_{s}x\right) \psi _{p}\left(
x\right) .$ Here $\psi _{p}$ and $\psi _{p}^{\prime }$ are the wavefunctions
before and after a Galilean boost to the superfluid's frame of reference.

When (2.37) is substituted into (2.36), the result is 
\begin{equation}
\hspace{-1.5cm}
S_{a}=-\int d^{4}x~\psi _{a}^{\dagger }
\left[ \left( \frac{1}{2}mv^{\mu
}v_{\mu }-\frac{1}{2m}\partial ^{\mu }\partial _{\mu }-\mu _{ext}\right)
-i\left( \frac{1}{2}\partial ^{\mu }v_{\mu }+v^{\mu }\partial _{\mu }\right)
\right] \psi _{a}.
\end{equation}
If $n_{s}$ and $v_{\mu }$ are slowly varying, so that $P_{ext}$
and $\partial ^{\mu }v_{\mu }$ can be neglected, (2.30) yields the
simplification 
\begin{equation}
S_{a}=\int d^{4}x~\psi _{a}^{\dagger }\left( \frac{1}{2m}\partial ^{\mu
}\partial _{\mu }+iv_{\alpha }^{\mu }\sigma ^{\alpha }\partial _{\mu
}\right) \psi _{a}.
\end{equation}
In the present theory, the gravitational vierbein is interpreted as the
``superfluid velocity'' associated with the GUT condensate $\Psi _{cond}$: 
\begin{equation}
e_{\alpha }^{\mu }=v_{\alpha }^{\mu }.
\end{equation}
Bosonic fields are conventionally represented as dimension $1$ (rather than
dimension $3/2$) operators, so let us define 
\begin{equation}
\phi _{b}=\psi _{b}/\left( 2m\right) ^{1/2}.
\end{equation}
Then the action for a free bosonic field is 
\begin{equation}
S_{b}=\int d^{4}x~\phi _{b}^{\dagger }\left( \partial ^{\mu }\partial
_{\mu }+2mie_{\alpha }^{\mu }\sigma ^{\alpha }\partial _{\mu }\right) \phi
_{b}
\end{equation}
with 
\begin{equation}
S_{b}~\rightarrow ~\int d^{4}x~\phi _{b}^{\dagger }\partial ^{\mu }\partial
_{\mu }\phi _{b}\quad \mbox{as}\;p_{\mu }\rightarrow \infty
\end{equation}
for a plane-wave state $\phi _{b}\propto \exp \left( ip_{\mu }x^{\mu
}\right) $. The usual form of the action for a massless and noninteracting
bosonic field is thus regained at high energy.

For a free fermionic field, on the other hand, the action is 
\begin{equation}
S_{f}=\int d^{4}x~\psi _{f}^{\dagger }\left( \frac{1}{2m}\partial ^{\mu
}\partial _{\mu }+ie_{\alpha }^{\mu }\sigma ^{\alpha }\partial _{\mu
}\right) \psi _{f}
\end{equation}
with 
\begin{equation}
S_{f}~\rightarrow ~\int d^{4}x~\psi _{f}^{\dagger }\,ie_{\alpha }^{\mu
}\sigma ^{\alpha }\,\partial _{\mu }\psi _{f}\quad \mbox{as}\;p_{\mu
}\rightarrow 0
\end{equation}
so the usual form of the action for a massless and noninteracting fermionic
field is regained at low energy. To be more specific, the standard fermionic
action is regained when 
\begin{equation}
p^{\mu }\ll mv_{\alpha }^{\mu }
\end{equation}
with $m\sim m_{P}$.

\section{Origin of Gauge Fields}

Let us now relax assumption (2.19) and allow $U_{int}$ to vary with
the external coordinates $x^{\mu }$. It is convenient to write 
\begin{equation}
\hspace{-1.5cm}
\Psi _{int}\left( x^{m}\right) =\widetilde{U}_{int}
\left( x^{\mu
},x^{m}\right) \bar{\Psi }_{int}\left( x^{m}\right) =\widetilde{U}
_{int}\left( x^{\mu },x^{m}\right) \bar{U}_{int}\,\left( x^{m}\right)
n_{int}^{1/2}\left( x^{m}\right) \eta _{int}
\end{equation}
where $n_{int}\left( x^{m}\right) =\bar{\Psi }_{int}^{\dagger }\left(
x^{m}\right) \bar{\Psi }_{int}\left( x^{m}\right) $ and $\bar{\Psi }_{int}$ 
still satisfies the internal equation of motion 
\begin{equation}
\left( -\frac{1}{2m}\partial ^{m}\partial _{m}-\mu _{int}+V_{vac}\right) 
\bar{\Psi }_{int}\left( x^{m}\right) =0.
\end{equation}
This is a nonlinear equation because $V_{vac}$ is largely determined by 
$n_{int}$.

The internal basis functions satisfy (2.35) with $\varepsilon _{r}=0$: 
\begin{equation}
\left( -\frac{1}{2m}\partial ^{m}\partial _{m}-\mu _{int}+V_{vac}\right)
\psi _{r}^{int}\left( x^{m}\right) =0.
\end{equation}
This is a linear equation because $V_{vac}\left( x^{m}\right) $ is now
regarded as a known function.

If the vacuum of the internal space had a trivial topology, the solutions to
(3.2) and (3.3) would be trivial, and the resulting universe would presumably 
not support nontrivial structures such as intelligent life. The full path 
integral involving (1.1) contains all configurations of the fields, however, 
including those with nontrivial topologies. In the present theory, the
``geography'' of the universe inhabited by human beings involves an internal
instanton in 
\begin{equation}
d=D-4
\end{equation}
dimensions which is analogous to a $U(1)$ vortex in $2$ dimensions or an 
$SU(2)$ instanton in $4$ Euclidean dimensions. The standard features of
four-dimensional physics -- including gauge symmetries and chiral fermions
-- arise from the presence of this instanton.

In the following, it is not necessary to have a detailed knowledge of the
internal instanton. The only property required is a $d$-dimensional
spherical symmetry for the internal condensate, and, as a result, for the 
functions $\widetilde{\psi }_{r}^{int}$ defined by 
\begin{equation}
\psi _{r}^{int}=\bar{U}_{int}\widetilde{\psi }_{r}^{int}.
\end{equation}
To be specific, it is required that 
\begin{equation}
K_{i}\;\widetilde{\psi }_{r}^{int}=0
\end{equation}
where 
\begin{equation}
K_{i}=K_{i}^{n}\partial _{n}
\end{equation}
is a Killing vector associated with the spherical symmetry of the internal
metric tensor $g_{mn}$ defined below. At a given point, the derivatives of
(3.7) involve only the $\left( d-1\right) $ angular coordinates, and not the
radial coordinate $r,$ so (3.6) states that $n_{int}$ and the 
$\widetilde{\psi }_{r}^{int}$ are functions only of $r$.

Although a detailed description is not necessary, it is worthwhile to
consider a concrete example, in which $V_{vac}=bn_{ext}n_{int}+V_{0}$ and 
$V_{0}$ is a constant. For clarity, we can start with a picture in which the
instanton occupies an unbounded volume, and then move to a physically more
acceptable description in which it is confined to a finite region $r<r_{0}$.
The finite instanton has finite action, and can be viewed as a ``spinning''
ball of condensate. The corresponding order parameter has a node at $r=r_{0}$, 
from which the condensate rises to become fully formed at large $r$. The
region $r<r_{0}$ corresponds to our physical universe, and the region 
$r>r_{0}$ is unobservable.

The same arguments that led to the external Bernoulli equation (2.25) also
yield an internal Bernoulli equation 
\begin{equation}
-\frac{1}{2m}n_{int}^{-1/2}\partial ^{m}\partial _{m}n_{int}^{1/2}+
\frac{1}{2}m\eta _{int}^{\dagger }v^{m}v_{m}\eta _{int}-\mu _{int}+V_{vac}=0.
\end{equation}
In our example, it is assumed that the instanton has the symmetry of a 
$\left( d-1\right) $-sphere, with 
\begin{equation}
\eta _{B}^{\dagger }v^{m}v_{m}\eta _{B}=\left( \bar{a}/mr\right) ^{2}
\end{equation}
\begin{equation}
\partial ^{m}\partial _{m}n_{int}^{1/2}=\frac{1}{r^{d-1}}\frac{d}{dr}\left(
r^{d-1}\frac{d}{dr}n_{int}^{1/2}\right) .
\end{equation}
Then (3.8) can be rewritten as 
\begin{equation}
-\frac{1}{\rho ^{d^{\prime }}}\frac{d}{d\rho }\left( \rho ^{d^{\prime }}
\frac{df}{d\rho }\right) +\frac{\bar{a}^{2}}{\rho ^{2}}f+f^{3}-f=0
\end{equation}
where $d^{\prime }=d-1$, $\rho =r/\xi _{int}$, 
and $f=n_{int}^{1/2}/\bar{n}_{int}^{1/2}$, with 
$\xi _{int}=\left( 2m\mu _{int}^{\prime }\right) ^{-1/2}$, $\mu
_{int}^{\prime }=\mu _{int}-V_{0},$ and $\bar{n}_{int}=\mu _{int}^{\prime
}/bn_{ext}$. The asymptotic solutions to (3.11) are 
\begin{equation}
f\propto \rho ^{n}\quad \mbox{as }\rho \rightarrow 0
\end{equation}
\begin{equation}
f=1-\bar{a}^{2}/2\rho ^{2}\quad \mbox{as }\rho \rightarrow \infty
\end{equation}
where 
\begin{equation}
n=\frac{1}{2}\left[ \sqrt{\left( d-2\right) ^{2}+4\bar{a}^{2}}-\left(
d-2\right) \right]
\end{equation}
so that 
\begin{equation}
n=1\quad \mbox{if }\bar{a}^{2}=d-1.
\end{equation}
It is easy to show that (3.15) holds for a minimal vortex in two dimensions
or a minimal $SU(2)$ instanton in four dimensions.

Since the volume element is proportional to $\rho ^{d-1}d\rho $ and $1-f^{2}$
is proportional to $\rho ^{-2}$ as $\rho \rightarrow \infty $, the above
solution has infinite action. However, we can obtain a solution with finite
action by requiring that 
\begin{eqnarray}
\Psi _{int} &=&R\left( r\right) \,\bar{n}_{int}^{1/2}\,U_{int}\,\eta _{int}\quad
,\quad \rho <\rho _{0} \\
\Psi _{int} &=&0\quad ,\quad \rho =\rho _{0} \\
\Psi _{int} &=&\bar{R}\left( r\right) \,\eta _{int}\quad ,\quad \rho >\rho _{0}
\end{eqnarray}
so that the instanton is confined to the region inside a radius $\rho _{0}$
which is determined by the boundary conditions below. Then (3.11) is
replaced by 
\begin{eqnarray}
-\frac{1}{\rho ^{d^{\prime }}}\frac{d}{d\rho }\left( \rho ^{d^{\prime }}
\frac{dR}{d\rho }\right) +\frac{\bar{a}^{2}}{\rho ^{2}}R+R^{3}-R &=&0\quad
,\quad \rho <\rho _{0} \\
-\frac{1}{2m}\frac{1}{r^{d^{\prime }}}\frac{d}{dr}\left( r^{d^{\prime }}
\frac{d\bar{R}}{dr}\right) +bn_{ext}\bar{R}^{3}-\mu \bar{R} &=&0\quad ,\quad
\rho >\rho _{0}.
\end{eqnarray}
$R$ is required to satisfy (3.19) with the boundary condition $R\rightarrow
0+$ as $\rho \rightarrow 0$. $\bar{R}$ is required to satisfy (3.20) with
the boundary condition $\bar{R}\rightarrow -\left( \mu /bn_{ext}\right) ^{1/2}$ 
as $r\rightarrow \infty $ (and with $\partial \Psi _{int}/\partial r$
continuous at $\rho =\rho _{0}$). In the following, we will be concerned
only with the physical region $\rho <\rho _{0}$, and the integrals are over
only this region; e.g., 
\begin{equation}
V_{int}=\int d^{d}x=\int_{\rho <\rho _{0}}d^{d}x.
\end{equation}

The above treatment assumes that the second-order equations (3.19) and
(3.20) are exact. However, in a 
picture that will be presented elsewhere~\cite{allen2}, the continuum 
approximation is not perfect, and as a result higher derivatives 
can be significant near the Planck scale. 
For an $n$th order differential equation, 
we have the freedom to impose $n$ boundary conditions. This fact
makes it possible to satisfy (3.16)-(3.17) for various values of $\rho _{0}$, 
so that the volume $V_{int}$ of the internal space is largely arbitrary.
As in other Kaluza-Klein theories,  $V_{int}$ determines the strength
of gravitational and gauge interactions, so the arbitrariness of $V_{int}$
has obvious anthropic implications.

The vierbein $e_{\alpha }^{\mu }$ of external spacetime was defined in
(2.40). It is convenient to define the remaining components of the vielbein
in a slightly different way, by representing $mv_{M}$ in terms of a set of
matrices $\sigma ^{A}$, 
\begin{equation}
v_{M}=v_{M A}\sigma ^{A}=v_{M \alpha }\sigma ^{\alpha }+v_{M c}\sigma ^{c},
\end{equation}
and letting 
\begin{equation}
e_{M c}=-v_{M c}\quad ,\quad M=0,1,...,D-1\quad ,\quad c\geq 4.
\end{equation}
(The $\sigma ^{\alpha }$ are associated with $U_{ext}$, and the $\sigma ^{c}$
with $U_{int}$. Since (2.16) implies that $v_{m \alpha }=0$, all the
nonzero $e_{M A}$ have now been specified.) When (2.19) holds, the only
nonzero components of the metric tensor are 
\begin{equation}
g^{\mu \nu }=\eta ^{\alpha \beta }e_{\alpha }^{\mu }e_{\beta }^{\nu }.
\end{equation}
and 
\begin{equation}
g_{mn}=e_{m c}e_{n c}
\end{equation}
which are respectively associated with external spacetime and the internal
space. More generally, however, $mv_{\mu }$ contains a contribution 
\begin{equation}
mv_{\mu c}\sigma ^{c}=-i\widetilde{U}_{int}^{-1}\left( x^{\mu
},x^{m}\right) \partial _{\mu }\widetilde{U}_{int}\left( x^{\mu
},x^{m}\right)
\end{equation}
so that $e_{\mu c}$ is nonzero and the metric tensor has off-diagonal
components 
\begin{equation}
g_{\mu m}=e_{\mu c}e_{m c}.
\end{equation}
In the present theory, just as in classic Kaluza-Klein theories, it is
appropriate to write 
\begin{equation}
e_{\mu c}=A_{\mu }^{i}K_{i}^{n}v_{n c}\quad ,\quad g_{\mu m}=
A_{\mu}^{i}K_{i}^{n}g_{mn}
\end{equation}
or, for later convenience, 
\begin{equation}
mv_{\mu c}\sigma ^{c} = -A_{\mu }^{i}\sigma _{i} 
\end{equation}
\begin{equation}
\sigma _{i} = mK_{i}^{n}v_{n c}\sigma ^{c}.
\end{equation}

For simplicity of notation, let 
\begin{equation}
\left\langle r|Q|s\right\rangle =\int d^{d}x\,\psi _{r}^{int\dagger }Q\psi
_{s}^{int}\quad \mbox{with}\quad \left\langle r|s\right\rangle =\delta _{rs}
\end{equation}
for any operator $Q$, so that (3.5)-(3.7) and (2.16) give 
\begin{equation}
\left\langle r|\left( -iK_{i}\right) |s\right\rangle =\left\langle r|\left(
-iK_{i}^{n}\right) \left( imv_{n}\right) |s\right\rangle =\left\langle
r|\sigma _{i}|s\right\rangle .
\end{equation}
With the definition 
\begin{equation}
t_{i}^{rs}=\left\langle r|\left( -iK_{i}\right) |s\right\rangle
\end{equation}
we then have 
\begin{equation}
\left\langle r|\sigma _{i}|s\right\rangle =t_{i}^{rs}.
\end{equation}
The Killing vectors have an algebra 
\begin{equation}
K_{i}K_{j}-K_{j}K_{i}=-c_{ij}^{k}K_{k}
\end{equation}
or 
\begin{equation}
\left( -iK_{i}\right) \left( -iK_{j}\right) -\left( -iK_{j}\right) \left(
-iK_{i}\right) =ic_{ij}^{k}\left( -iK_{k}\right)
\end{equation}
so the same is true of the matrices $t_{i}^{rs}$: 
\begin{equation}
t_{i}t_{j}-t_{j}t_{i}=ic_{ij}^{k}t_{k}.
\end{equation}

With the more general version of (2.33) and (2.37), 
\begin{equation}
\Psi _{a}\left( x^{\mu },x^{m}\right) =U_{ext}\left( x^{\mu }\right) 
\widetilde{U}_{int}\left( x^{\mu },x^{m}\right) \psi _{a}^{r}\left( x^{\mu
}\right) \psi _{r}^{int}\left( x^{m}\right) ,
\end{equation}
we have 
\begin{equation}
\partial _{\mu }\Psi _{a}=U_{ext}\left( x^{\mu }\right) \widetilde{U}
_{int}\left( x^{\mu },x^{m}\right) \left( \partial _{\mu }+imv_{\mu \alpha
}\sigma ^{\alpha }+imv_{\mu c}\sigma ^{c}\right) \psi _{a}^{r}\psi _{r}^{int}
\end{equation}
and 
\begin{eqnarray} 
\hspace{-2.0cm}
&{}&\int d^{d}x \, \Psi _{a}^{\dagger }\, 
\partial ^{\mu }\partial _{\mu } \, \Psi _{a}  \nonumber\\
\hspace{-2.0cm}
&=&\int d^{d}x \, \psi _{r}^{int \dagger }\psi _{a}^{r \dagger}\eta ^{\mu \nu }
\left( \partial _{\mu }+imv_{\mu \alpha }\sigma ^{\alpha}
+imv_{\mu c}\sigma ^{c}\right) \left( \partial _{\nu }+imv_{\nu \beta}
\sigma ^{\beta }+imv_{\nu d}\sigma ^{d}\right) \psi _{a}^{s}\psi
_{s}^{int}
\nonumber \\
\hspace{-2.0cm}
&=&\psi _{a}^{r \dagger}\,\eta ^{\mu \nu }\langle r|
\left( \partial _{\mu }+imv_{\mu \alpha}
\sigma ^{\alpha }+imv_{\mu c}\sigma ^{c}\right)
\,\sum_{t}|t\rangle \langle t|\,\left( \partial _{\nu }+imv_{\nu \beta}
\sigma ^{\beta }+imv_{\nu d}\sigma ^{d}\right) |s\rangle \,\psi
_{a}^{s} 
\nonumber \\
\hspace{-2.0cm}
&=&\psi _{a}^{r \dagger }\,\eta ^{\mu \nu }\left[ \delta _{rt}
\left( \partial _{\mu}
+imv_{\mu \alpha }\sigma ^{\alpha }\right) -iA_{\mu}^{i}
t_{i}^{rt}\right] \,\left[ \delta _{ts}\left( \partial _{\nu}+imv_{\nu \beta}
\sigma ^{\beta }\right) -iA_{\nu }^{j}t_{j}^{ts}\right] \,\psi
_{a}^{s} 
\nonumber \\
\hspace{-2.0cm}
&=&\psi _{a }^{\dagger }\,\eta ^{\mu \nu }
\left[ \left( \partial _{\mu }-iA_{\mu}^{i}
t_{i}\right) +imv_{\mu \alpha }\sigma ^{\alpha }\right] \,\left[ \left(
\partial _{\nu }-iA_{\nu }^{j}t_{j}\right) +imv_{\nu \beta }\sigma
^{\beta }\right] \,\psi _{a}
\end{eqnarray}
\newline
where (2.34), (3.29), and (3.34) have been used. The action (2.32)
then becomes
\begin{equation}
\hspace{-2.0cm}
S_{a}=\int d^{4}x\,\psi _{a}{}^{\dagger }\left( \frac{1}{2m}D^{\mu }D_{\mu }+
\frac{1}{2}iv_{\alpha }^{\mu }\sigma ^{\alpha }D_{\mu }+\frac{1}{2}D_{\mu
}iv_{\alpha }^{\mu }\sigma ^{\alpha }-\frac{1}{2}mv^{\alpha \mu }v_{\mu
}^{\alpha }+\mu _{ext}\right) \psi _{a}
\end{equation}
after (2.35) is used, where 
\begin{equation}
D_{\mu }=\partial _{\mu }-iA_{\mu }^{i}t_{i}.
\end{equation}
With the approximations above (2.39), (2.30) and (2.40) imply that 
\begin{equation}
S_{a}=\int d^{4}x\,\psi _{a}{}^{\dagger }\left( \frac{1}{2m}D^{\mu }D_{\mu
}+ie_{\alpha }^{\mu }\sigma ^{\alpha }D_{\mu }\right) \psi _{a}.
\end{equation}
This is in fact the generalization of (2.39) when the ``internal order
parameter'' is permitted to vary as a function of the external coordinates 
$x^{\mu }$.

As in Ref. 1, let us postulate a cosmological model in which 
\begin{equation}
e_{\alpha }^{\mu }=\lambda \delta _{\alpha }^{\mu }\equiv \widetilde{e}
_{\alpha }^{\mu }.
\end{equation}
In this case (3.43) can be rewritten as 
\begin{equation}
S_{a}=\int d^{4}x\,\widetilde{g}\,\bar{\psi }_{a}{}^{\dagger }\left( 
\bar{m}^{-1}\widetilde{g}^{\mu \nu }D_{\mu }D_{\nu }+ie_{\alpha }^{\mu
}\sigma ^{\alpha }D_{\mu }\right) \bar{\psi }_{a}
\end{equation}
where 
\begin{eqnarray}
\widetilde{g}^{\mu \nu } &\equiv &\eta ^{\alpha \beta }\widetilde{e}_{\alpha
}^{\mu }\widetilde{e}_{\beta }^{\nu }\quad ,\quad \bar{m}=2\lambda ^{2}m
\\
\widetilde{g} &=&\left( -\det \widetilde{g}_{\mu \nu }\right) ^{1/2}=\lambda
^{-4}\quad ,\quad \bar{\psi }_{a}=\lambda^{2}\psi _{a}.
\end{eqnarray}
(The tilde is a reminder that the above form is not general, and that 
$\widetilde{g}^{\mu \nu }$ is not a dynamical quantity.) In a locally
inertial coordinate system with $e_{\alpha }^{\mu }=\delta _{\alpha }^{\mu }$
, this becomes 
\begin{equation}
S_{a}=\int d^{4}x\,\,\psi _{a}{}^{\dagger }\left( \bar{m}^{-1}\eta
^{\mu \nu }D_{\mu }D_{\nu }+i\sigma ^{\mu }D_{\mu }\right) \psi _{a}
\end{equation}
where the bar has been removed from $\psi _{a}$ for simplicity, so 
the fermionic and bosonic actions are respectively 
\begin{equation}
S_{f}=\int d^{4}x\,\,\psi _{f}{}^{\dagger }\left( \bar{m}^{-1}\eta
^{\mu \nu }D_{\mu }D_{\nu }+i\sigma ^{\mu }D_{\mu }\right) \psi _{f}
\end{equation}
and 
\begin{equation}
S_{b}=\int d^{4}x\,\phi _{b}{}^{\dagger }\left( \eta ^{\mu \nu }D_{\mu
}D_{\nu }+i\bar{m}\sigma ^{\mu }D_{\mu }\right) \phi _{b}
\end{equation}
where now 
\begin{equation}
\phi _{b}=\psi _{b}/\bar{m}^{1/2}.
\end{equation}
Again, one regains the usual bosonic action at high energy, 
\begin{equation}
S_{b}~\rightarrow ~\int d^{4}x~\phi _{b}^{\dagger }\eta ^{\mu \nu }D_{\mu
}D_{\nu }\phi _{b}\quad \mbox{for}\;p^{\mu }\gg \bar{m},
\end{equation}
and the usual fermionic action at low energy, 
\begin{equation}
S_{f}~\rightarrow ~\int d^{4}x~\psi _{f}^{\dagger }\,i\sigma ^{\mu }\,D_{\mu
}\psi _{f}\quad \mbox{for}\;p^{\mu }\ll \bar{m},
\end{equation}
where the expressions now include gauge couplings and are written in a
locally inertial coordinate system.

\pagebreak

Recall that the initial gauge group is the same as the group of rotations in
the internal space -- e.g., $SO(10)$ for $d=10$. The generators $t_{i}$
correspond to a reducible representation of this group, composed of some set
of irreducible representations that are left unspecified in the present
paper, although it is clear that one can place the three generations of 
Standard Model fermions in three spinorial ${\bf 16}$ representations. Each 
field will necessarily have a superpartner with the same quantum numbers,
just as in standard supersymmetry~\cite{kane-review,kane}. We leave the 
phenomenology of these fields for future work.

Notice that the deviations from standard physics in (3.49) and (3.50) are
predicted only for (i)~fermions at very high energy and (ii)~fundamental
bosons which have not yet been observed. Notice also that the present theory
preserves both gauge invariance and many features of Lorentz invariance,
including rotational invariance and the requirement that all massless 
particles travel with the same speed $c=1$ in a locally inertial coordinate 
system. (This last feature follows from (4.18)-(4.21).) It appears 
that the present theory is in agreement with even the most sensitive tests
of Lorentz invariance that are currently available~\cite{kostelecky}. 
Furthermore, issues like causality and logical consistency 
can ultimately be resolved by returning to the original action (1.1).

\section{Consistency of Canonical Quantization}

Let us now consider whether the present theory permits a consistent
extension of standard field theory~[16-24]. This is not a trivial issue
because, as mentioned above, the fermionic action (3.49) is Lorentz invariant
only at low energy ($p^{\mu }\ll \bar{m}$), and the bosonic action (3.50)
has its usual form only at high energy ($p^{\mu }\gg \bar{m}$). 

We will, in fact, encounter a difficulty which is essentially the same as
that encountered in covariant quantization of the electromagnetic field~\cite
{mandl}. Let $\zeta _{p\lambda }$ be the norm of a one-particle state 
$|1_{p\lambda }\rangle :$ 
\begin{equation}
\left\langle 1_{p\lambda }|1_{p\lambda }\right\rangle =\zeta _{p\lambda}
\quad , \quad |1_{p\lambda }\rangle =a_{p\lambda }^{\dagger}|0\rangle 
\end{equation}
where $p$ is the momentum and $\lambda $ is the index defined below in (4.24). 
As in the case of the electromagnetic field, the quantization condition (4.30) 
or (4.48) will imply that there are intermediate states with negative norm. Even 
though such states can be consistently treated with the formalism of Gupta 
and Bleuler~\cite{mandl,jauch,kallen}, they are not physical, so it is
necessary to require that 
\begin{eqnarray}
a_{p\lambda }\left| phys\right\rangle  &=&0\quad \mbox{if}\quad \omega
_{p\lambda }>0\quad \mbox{and}\quad \zeta _{p\lambda }<0 \\
b_{p\lambda }\left| phys\right\rangle  &=&0\quad \mbox{if}\quad \omega
_{p\lambda }<0\quad \mbox{and}\quad \zeta _{p\lambda }<0
\end{eqnarray}
where $a_{p\lambda }$ and $b_{p\lambda }$ are the particle and antiparticle
destruction operators introduced below, and $\left| phys\right\rangle $
represents any physical state. (For the electromagnetic field, one can
choose a gauge such that this condition is satisfied separately for 
all unphysical photons, although in a general Lorentz gauge it can be
relaxed to $\left( a_{p3}-a_{p0}\right) \left| phys\right\rangle =0$, because
the contributions of longitudinal and scalar photons then cancel.) For
fermions in the present context, the condition for a single-particle state
with positive norm will turn out to be 
\begin{equation}
1+2\omega _{n}/\bar{m}>0.
\end{equation}
(See (4.33).) According to (4.18)-(4.21), there is always one function of
the form $u_{p}\exp i\vec{p}\cdot \vec{x}$ satisfying this requirement, with 
\begin{equation}
\lambda =1
\end{equation}
where $u_{p}$ is the right-handed 2-component spinor defined below in
(4.22). There is also one positive-norm function of the form 
$v_{p}\exp i\vec{p}\cdot \vec{x}$, with 
\begin{equation}
\lambda =3\;\mbox{ for }\;\left| \vec{p}\right| <\bar{m}/2\quad , 
\quad \lambda =4\;\mbox{ for }\;\left| \vec{p}\right| >\bar{m}/2
\end{equation}
where $v_{p}$ is the left-handed 2-component spinor defined below in (4.23).
For bosons, on the other hand, the states  with positive norm must satisfy 
\begin{equation}
1+\bar{m}/2\omega _{n}>0.
\end{equation}
(See (4.50).) There are always two functions of the form 
$u_{p}\exp i\vec{p}\cdot \vec{x}$ satisfying this condition, with 
\begin{equation}
\lambda =1\mbox{ or }2.
\end{equation}
There is also at least one positive-norm function of the form 
$v_{p}\exp i\vec{p}\cdot \vec{x}$, with 
\begin{equation}
\hspace{-0.95in}
\lambda =4\; \mbox{for}\; \left| \vec{p}\right| <\bar{m}/2 , \quad 
\lambda =3\ \; \mbox{for} \; \bar{m}/2<\left| \vec{p}\right|<\bar{m} , 
\quad  \lambda =3 \; \mbox{or} \; 4 \; \mbox{for} \; 
\left| \vec{p}\right|>\bar{m}.
\end{equation}
In summary, $\zeta _{p\lambda }$ is $>0$ for one-particle states satisfying
(4.4)-(4.6) for fermions and (4.7)-(4.9) for bosons, with 
$\zeta _{p\lambda }<0$ otherwise.

Let $\psi $ and $\phi $ represent 2-component, complex fermionic and bosonic
fields. In the case of fermions, and with gauge fields omitted, (3.49) gives
\begin{eqnarray}
{\cal L}_{\psi } &=&-\,\bar{m}^{-1}\eta ^{\mu \nu }\partial _{\mu }
\psi ^{\dagger}\partial _{\nu }\psi 
+\frac{1}{2} \left( i\psi ^{\dagger }\sigma ^{\mu }
\partial _{\mu }\psi + h.c. \right)  \\
&=&\bar{m}^{-1}\left( \dot {\psi }^{\dagger }\dot
{\psi }-\,\partial ^{k}\psi ^{\dagger }\partial _{k}\psi \right) +\frac{1}{2}
\left( i\psi ^{\dagger }\dot{\psi }+\,i\psi ^{\dagger }\sigma
^{k}\partial _{k}\psi + h.c. \right) 
\end{eqnarray}
where $\dot{\psi }=\partial _{0}\psi$ and ``$h.c.$'' means 
``Hermitian conjugate''. The canonical momenta
are (in a convenient but slightly unconventional notation) 
\begin{eqnarray}
\pi _{\psi }^{\dagger } &=& \frac{\partial {\cal L}_{\psi }}{\partial 
\dot{\psi }} = \;\bar{m}^{-1}\dot{\psi }^{\dagger }
+\frac{1}{2}i\psi ^{\dagger }  \\
\pi _{\psi } &=& \frac{\partial {\cal L}_{\psi }}{\partial 
\dot{ \psi }^{\dagger }}=\bar{m}^{-1}\dot{\psi }-\frac{1}{2}i\psi 
\end{eqnarray}
and the Hamiltonian density is 
\begin{eqnarray}
{\cal H}_{\psi } &=&\pi _{\psi }^{\dagger }\dot{\psi }+
\dot {\psi }^{\dagger }\pi _{\psi }-{\cal L}_{\psi } \\
&=&\bar{m}^{-1}\left( \dot{\psi }^{\dagger }\dot
{\psi }+\partial ^{k}\psi ^{\dagger }\partial _{k}\psi \right) -\frac{1}{2}
\left( i\psi ^{\dagger }\sigma ^{k}\partial _{k}\psi + h.c. \right) .
\end{eqnarray}
From (4.10) we obtain the equation of motion 
\begin{equation}
\bar{m}^{-1}\,\eta ^{\mu \nu }\partial _{\mu }\partial _{\nu }\psi +i\sigma
^{\mu }\partial _{\mu }\psi =0.
\end{equation}
Let $a_{n}\psi _{n}$ be a solution to this equation. Then we can write 
\begin{equation}
\psi =\sum_{n}\,a_{n}\psi _{n} .
\end{equation}
For each 3-momentum $\vec{p}$, there are four solutions to (4.16): 
\begin{eqnarray}
\psi _{p1} &=&A_{p1}\;u_{p}\;e^{i\vec{p}\cdot \vec{x}}\quad
,\;a_{p1}=e^{-i\omega _{p1}x^{0}}a_{p1}\left( 0\right) \quad ,\;\omega
_{p1}=\left| \vec{p}\right|  \\
\psi _{p2} &=&A_{p2}\;u_{p}\;e^{i\vec{p}\cdot \vec{x}}\quad
,\;a_{p2}=e^{-i\omega _{p2}x^{0}}a_{p2}\left( 0\right) \quad ,\;\omega
_{p2}=-\bar{m}-\left| \vec{p}\right|  \\
\psi _{p3} &=&A_{p3}\;v_{p}\;e^{i\vec{p}\cdot \vec{x}}\quad
,\;a_{p3}=e^{-i\omega _{p3}x^{0}}a_{p3}\left( 0\right) \quad ,\;\omega
_{p3}=-\left| \vec{p}\right|  \\
\psi _{p4} &=&A_{p4}\;v_{p}\;e^{i\vec{p}\cdot \vec{x}}\quad
,\;a_{p4}=e^{-i\omega _{p4}x^{0}}a_{p4}\left( 0\right) \quad ,\;\omega
_{p4}=-\bar{m}+\left| \vec{p}\right| 
\end{eqnarray}
where 
\begin{eqnarray}
\vec{\sigma}\cdot \vec{p}\,u_{p} &=&+\left| \vec{p}\right| u_{p} \\
\vec{\sigma}\cdot \vec{p}\,v_{p} &=&-\left| \vec{p}\right| v_{p}
\end{eqnarray}
\begin{equation}
n \leftrightarrow \vec{p},\lambda \quad \mbox{with }\lambda =1,2,3,4
\end{equation}
and the $A_{p\lambda }$ are normalization constants specified below. We can
choose 
\begin{eqnarray}
u_{p}^{\dagger }u_{p}=v_{p}^{\dagger }v_{p} &=&1\quad ,\quad u_{p}^{\dagger
}v_{p}=v_{p}^{\dagger }u_{p}=0 \\
u_{p}u_{p}^{\dagger }+v_{p}v_{p}^{\dagger } &=&{\bf 1}
\end{eqnarray}
where ${\bf 1}$ is the $2\times 2$ identity matrix. Since 
\begin{equation}
\dot{a}_{n}=-i\omega _{n}a_{n}\quad ,\quad \dot{a}
_{n}^{\dagger }=i\omega _{n}a_{n}^{\dagger }
\end{equation}
(4.12) and (4.13) give 
\begin{eqnarray}
\pi _{\psi }^{\dagger } &=& \frac{1}{2}i\sum_{n}\,\left( 1+
2\omega _{n}/\bar{m}\right) a_{n}^{\dagger }\psi _{n}^{\dagger }  \\
\pi _{\psi } &=& -\frac{1}{2}i\sum_{n}\,\left( 1+2\omega _{n}/\bar{m}\right)
a_{n}\psi _{n}.
\end{eqnarray}
We quantize by interpreting $\psi $ and $\pi ^{\dagger }$ as operators, and
requiring that 
\begin{equation}
\left[ \psi \left( \vec{x},x^{0}\right) ,\pi _{\psi }^{\dagger }\left( \vec{x}
\,^{\prime },x^{0}\right) \right] _{+}=i\delta \left( \vec{x}
-\vec{x}\,^{\prime}\right) {\bf 1}
\end{equation}
or more explicitly 
\begin{equation}
\left[ \psi _{\alpha }\left( \vec{x},x^{0}\right) ,\pi _{\psi \beta
}^{\dagger }\left( \vec{x}\,^{\prime },x^{0}\right) \right] _{+}=i\delta
\left( \vec{x}-\vec{x}\,^{\prime }\right) \delta _{\alpha \beta }
\end{equation}
where $\alpha $ and $\beta $ label the two components of $\psi $ and $\pi 
_{\psi }^{\dagger }$, with $\left[ X,Y\right] _{\pm }=XY\pm YX$. This
requirement will be satisfied if 
\begin{eqnarray}
\left[ a_{n},a_{m}^{\dagger }\right] _{+} =\delta_{nm}\,\zeta_{n}^{f} \\
\zeta_{n}^{f} = sgn\left( 1+2\omega _{n}/\bar{m}\right) 
\end{eqnarray}
\begin{equation}
\left[ a_{n},a_{m}\right] _{+} = \left[ a_{n}^{\dagger },a_{m}^{\dagger
}\right] _{+}=0 
\end{equation}
\begin{equation}
A_{n}^{*}A_{n}=V^{-1}\left| 1+2\omega _{n}/\bar{m}\right|^{-1}
\end{equation}
where $V$ is the normalization volume, since
\begin{eqnarray}
\hspace{-1in}\lefteqn{\frac{1}{2}\sum_{n}\left|1+ 2\omega _{n}/\bar{m}
\right| \psi _{n}\left(
\vec{x} \right) \psi _{n}^{\dagger }\left( \vec
{x}\,^{\prime } \right) 
=\frac{1}{2}\sum_{\vec{p}\;\lambda =1,2}
\left|1+ 2\omega _{p\lambda }/\bar{m}\right|
A_{p\lambda }A_{p\lambda }^{*}u_{p}u_{p}^{\dagger }
e^{i\vec{p}\cdot \left( \vec{x}-
\vec{x}\,^{\prime }\right) }}
\hspace{1.35in}  \nonumber \\
&&\hspace{1.4in}+\frac{1}{2}\sum_{\vec{p}\;\lambda =3,4} \left|1+
2\omega _{p\lambda }/\bar{m}\right| A_{p\lambda }A_{p\lambda }^{*}
v_{p}v_{p}^{\dagger }e^{i\vec{p}\cdot \left( \vec{x}-
\vec{x}\,^{\prime }\right) } \hspace{-1in}   \nonumber \\
&=&V^{-1}\sum_{\vec{p}}e^{i\vec{p}\cdot \left(
\vec{x}-\vec{x}\,^{\prime }\right) }\left(
u_{p}u_{p}^{\dagger }+v_{p}v_{p}^{\dagger }\right)  \\
&=&\delta \left( \vec{x}-\vec{x}\,^{\prime }\right){\bf 1}.
\end{eqnarray}
The equation of motion (4.16) implies that 
${\cal L}_{\psi }$ has the form $\partial_{\mu}{\cal M}^{\mu}$, 
so the Hamiltonian density of (4.14) is 
\begin{equation}
{\cal H}_{\psi }=\pi _{\psi }^{\dagger }\dot{\psi }+
\dot {\psi }^{\dagger }\pi _{\psi } - \partial_{\mu}{\cal M}^{\mu} .
\end{equation}
Suppose for the moment that we are interested in the time-averaged 
value $\bar {H}$ of the Hamiltonian. In this context the last term can be 
ignored, since it does not contribute to the four-dimensional integral 
when boundary terms are ignored. In any state, we then have
\begin{eqnarray}
\left\langle \bar {H}_{\psi }\right\rangle  
&=&\int d^{3}x\,\left\langle \bar{\mathcal H}
_{\psi }\right\rangle  \\
&=&\sum_{n}\,\omega _{n}\left( 1+2\omega _{n}/\bar{m}\right) \left\langle
a_{n}^{\dagger }a_{n}\right\rangle \int d^{3}x\,\psi _{n}^{\dagger }\psi _{n}
\\
&=&\sum_{n,\omega _{n}>0}\left\langle a_{n}^{\dagger }a_{n} \,\zeta_{n}^{f} 
\right\rangle \left| \omega _{n}\right| -\sum_{n,\omega _{n}<0}\left\langle
b_{n}b_{n}^{\dagger } \,\zeta_{n}^{f} \right\rangle \left| \omega _{n}\right|  
\\
&=&\sum_{n}\left\langle N_{n}^{f} \right\rangle 
\left| \omega _{n}\right| -\sum_{n,\omega _{n}<0}\left|
\omega _{n}\right| 
\end{eqnarray}
where 
\begin{eqnarray}
N_{n}^{f} &=& a_{n}^{\dagger }a_{n} \,\zeta_{n}^{f} 
\; ,\;\omega_{n}>0  \\
N_{n}^{f} &=& 
b_{n}^{\dagger}b_{n} \,\zeta_{n}^{f} \; ,\;\omega _{n}<0
\end{eqnarray}
\begin{equation}
b_{n}^{\dagger }=a_{n}\quad ,\;b_{n}=a_{n}^{\dagger }
\end{equation}
\begin{equation}
\left[ b_{n},b_{m}^{\dagger }\right] _{+}=\left[ a_{n},a_{m}^{\dagger
}\right] _{+}=\delta _{nm}\,\zeta _{n}^{f}.
\end{equation}

The above treatment can be repeated for the fundamental bosons described by
(3.50), with 
\begin{equation}
\psi \rightarrow \phi ,\;a_{n}\rightarrow c_{n},\;A_{n}\rightarrow
B_{n},\;b_{n}\rightarrow d_{n}
\end{equation}
\begin{equation}
\left[ \phi \left( \vec{x},x^{0}\right) ,\pi _{\phi }^{\dagger }\left( 
\vec{x}\,^{\prime },x^{0}\right) \right] _{-}=i\delta \left( \vec{x}-\vec{x}
\,^{\prime }\right) \mathbf{1}
\end{equation}
\begin{eqnarray}
\left[ c_{n},c_{m}^{\dagger }\right] _{-} = \delta _{nm}\,\zeta
_{n}^{b} \; \omega _{n}/\left| \omega _{n}\right| \\
\zeta_{n}^{b} = sgn\left( 1+\bar{m}/2\omega _{n}\right)  
\end{eqnarray}
\begin{equation}
\left[ c_{n},c_{m}\right] _{-} = \left[ c_{n}^{\dagger },c_{m}^{\dagger
}\right] _{-}=0
\end{equation}
\begin{equation}
\phi _{n}^{\dagger }\left( \vec{x}\right) \phi _{n}\left( \vec{x}\right)
=B_{n}^{*}B_{n}=\left( 2\left| \omega _{n}\right| V\right)
^{-1}\left| 1+\bar{m}/2\omega _{n}\right| ^{-1}
\end{equation}
\begin{eqnarray}
N_{n}^{b} &=& c_{n}^{\dagger }c_{n} \,\zeta_{n}^{b} 
\; ,\;\omega_{n}>0  \\
N_{n}^{b} &=& d_{n}^{\dagger }d_{n} \,\zeta_{n}^{b} 
\; ,\;\omega _{n}<0
\end{eqnarray}
\begin{equation}
d_{n}^{\dagger }=c_{n}\quad ,\;d_{n}=c_{n}^{\dagger }
\end{equation}
\begin{equation}
\left[ d_{n},d_{m}^{\dagger }\right] _{-}=-\left[ c_{n},c_{m}^{\dagger
}\right] _{-}=\delta _{nm}\,\zeta _{n}^{b}\quad ,\;\omega _{n}<0.
\end{equation}
\begin{eqnarray}
\left\langle \bar {H}_{\phi }\right\rangle  &=&\sum_{n,\omega _{n}>0}\left\langle
c_{n}^{\dagger }c_{n} \,\zeta_{n}^{b} \right\rangle 
\left| \omega _{n}\right| +\sum_{n,\omega_{n}<0}
\left\langle d_{n}d_{n}^{\dagger } \,\zeta_{n}^{b} \right\rangle 
\left| \omega_{n}\right|  \\
&=&\sum_{n} \left\langle N_{n}^{b} \right\rangle 
\left| \omega _{n}\right| +\sum_{n,\omega _{n}<0}\left| \omega _{n}\right| .
\end{eqnarray}

Returning to (4.15), one can use (4.18)-(4.23) to obtain the  
operator $H$ itself:
\begin{eqnarray}
H &=& H_{\psi} + H_{\phi} \\ 
H_{\psi} &=&\sum_{n} N_{n}^{f} \left| \omega _{n} \right| 
-\sum_{n,\omega _{n}<0} \left| \omega _{n} \right| \\ 
H_{\phi} &=& \sum_{n} N_{n}^{b} \left| \omega _{n} \right| 
+\sum_{n,\omega _{n}<0} \left| \omega _{n} \right| .
\end{eqnarray}
The total energy in any state is then 
\begin{equation}
\left\langle H\right\rangle =
\sum_{n} \left\langle N_{n}^{f} \right\rangle \left| \omega _{n}\right| +
\sum_{n} \left\langle N_{n}^{b} \right\rangle \left| \omega _{n}\right| .
\end{equation}
In particular, there is a cancellation of the bosonic and fermionic
contributions to the vacuum energy (before the initial supersymmetry of the
present theory is broken) just as in standard supersymmetry~\cite{kane}:
\begin{equation}
\left\langle 0 \left | H \right| 0 \right\rangle = 0.
\end{equation}
These results are not as trivial as they may seem, because the Lagrangians 
of (3.49) and (3.50) violate Lorentz invariance, and the $\omega_{n}$ 
are given by (4.18)-(4.21).

Other conserved quantities will also have their usual forms, because they
also involve products like $\pi _{\psi }^{\dagger }\partial _{\nu }\psi $ or 
$\pi _{\psi }^{\dagger }\Delta \psi $: According to Noether's theorem 
for a single field $\chi$, a conserved current has the form~\cite{peskin,mandl} 
\begin{equation}
j^{\mu }=\frac{\partial {\mathcal{L}}}{\partial \left( \partial _{\mu }\chi
\right) }\Delta \chi +a^{\nu } \left[ \frac{\partial {\mathcal{L}}}
{\partial \left( \partial _{\mu }\chi \right) }\partial _{\nu }\chi 
- {\cal L}\delta^{\mu}{}_{\nu} \right]
\end{equation}
so that 
\begin{equation}
j^{0}=\pi _{\chi }^{\dagger }\Delta \chi + a^{\nu} \left( \pi_{\chi}^{\dagger}
\partial_{\nu}\chi - {\cal L}\delta^{0}{}_{\nu} \right)
\end{equation}
where each $a^{\nu}$ represents an independent shift of coordinates 
and $\Delta \chi$ represents the effect of a rotation 
or gauge transformation. Using the corresponding result for the two 
fields $\psi$ and $\psi^{\dagger}$, or $\phi$ and $\phi^{\dagger}$, 
one obtains the usual expressions for the momenta, angular momenta,  
and charges.

The most dramatic prediction of this paper is that the bosons described 
by $\phi$ have an equation of motion which requires them to 
transform as spin 1/2 particles. Furthermore, there is a clear 
breaking of particle-antiparticle symmetry in the representation of 
$\phi$ even at low energy. Both of these features are associated with the 
fact that the present theory violates Lorentz invariance. The 
inapplicability of the usual spin-statistics and CPT 
theorems~[16,31-33] will be discussed elsewhere.

\section{Conclusion}

Let us now summarize some of the results of the preceding sections. 

The action (1.1) implies that a GUT-scale condensate (2.8) forms in
the very early universe. It is assumed that two topological defects are
``frozen into'' this condensate as it forms: a cosmological instanton, which
results in $U(1)\times SU(2)$ rotations of the external order parameter 
$\Psi _{ext}$, and an internal instanton, which results in rotations of the
internal order parameter $\Psi _{int}$. Since the other fermionic and
bosonic fields are born into this primordial condensate, it is natural to 
transform them to the frame of reference that rotates with it. In external
spacetime, this leads to an action for fermions which is Lorentz-invariant
at low energy (compared to an energy scale $\bar{m}$ which is
presumably well above $1$ TeV). The action for the initial fundamental bosons 
has exactly the same form as that for fermions, and is therefore quite
unconventional.

Both fermions and bosons are found to have standard couplings to the gauge
fields of an $SO(d)$ theory, where $d$ is the dimension of the space
containing the internal instanton. With $d=10$, we obtain an $SO(10)$
grand-unified theory, which naturally leads to neutrino masses,
coupling-constant unification, etc. It was also shown that the fermionic and
bosonic fields can be quantized with either a path-integral or canonical
description, even though their equations of motion are unconventional.

In this paper we did not attempt to develop a detailed phenomenological
picture. However, the forms (3.49) and (3.50) imply that there are 
testable extensions of standard physics for fermions at high 
energy and for fundamental bosons which have not yet been observed. In 
particular, fermions have an equation of motion that violates Lorentz 
invariance at high energy, while the bosons discussed in the previous 
section violate particle-antiparticle symmetry, and other features 
associated with Lorentz invariance, even at low energy. For example, 
these bosons have spin 1/2.

\end{document}